# Evolution equation of quantum tomograms for a driven oscillator in the case of the general linear quantization

*G.G. Amosov[1], V.I. Man'ko[2], Yu.N. Orlov[3]*

[1] *Moscow Institute of Physics and Technology, E-mail: gramos@mail.ru*
[2] *P.N. Lebedev Physical Institute, E-mail: manko@sci.lebedev.ru*
[3] *M.V. Keldysh Institute for Applied Mathematics. E-mail: orlmath@keldysh.ru*

**Abstract.**

The symlectic quantum tomography for the general linear quantization is introduced. Using the approach based upon the Wigner function techniques the evolution equation of quantum tomograms is derived for a parametric driven oscillator.

## 1. Introduction

In [1-2] the tomographic representation of quantum statistical operators and its evolution equation was introduced for the Weyl quantization. As an example of the exactly solvable problem to describe the evolution of a quantum system under the action of some external force one can take a driven oscillator for which the integrals of motion and the propagator were obtained in [3]. In [4] the dynamical equations for the driven oscillator were derived and solved in the tomographic representation associated with the Weyl quantization. Notice that the presence of the dynamical equation allows to rebuild the quantum tomogram from the uncomplete information [5]. Our goal is to construct the tomographic representation for the driven oscillator in the case of the general linear quantization and to derive the evolution equation for the corresponding statistical operators. The approach we investigate is closely related to star-product schemes and properties of the product kernels, see e.g.[6,7,8].

## 2. The general linear quantization of dynamical system

Let $q, p$ be the canonical coordinates in the phase space of classical mechanics $\Gamma$. Consider some function $A(q, p): \Gamma \to R$. For the sake of brevity we shall consider a one-dimensional case, i.e. $\Gamma = R \times R$. Let $\hat{A}$ be the hermitian operator acting in the Hilbert space $L_2(R)$ by the formula

$$\forall \psi \in L_2 \quad \hat{A}\psi(x) = \int \widetilde{A}(x,y)\psi(y)dy, \tag{1}$$

where $\widetilde{A}(x,y)$ is the density matrix of $\hat{A}$ in some basis of $L_2(R)$. The function $A(q,p)$ is said to be a classical symbol of the operator $\hat{A}$ under the quantization with the kernel $K(q,p|x,y)$ (or, for the sake of brevity, with the kernel $K$) if

$$\widetilde{A}(x,y) = \int dq dp A(q,p) K(q,p|x,y). \tag{2}$$

The correspondence between the operators and the symbols will be denoted as $A \leftrightarrow \hat{A}$. Let us consider the quatizations for which the following correspondencies hold:

$$q \leftrightarrow x, \quad p \leftrightarrow -i\frac{\partial}{\partial x}, \quad q^n \leftrightarrow (\hat{q})^n, \quad p^n \leftrightarrow (\hat{p})^n. \tag{3}$$

According to [9] the correspondence between the operators and the symbols are completely determined by the formlae describing the symbols of the operators $\hat{p}\hat{A}$, $\hat{A}\hat{p}$, $\hat{q}\hat{A}$, $\hat{A}\hat{q}$ from the symbol of the operator $\hat{A}$. The quantization is said to be linear if these formulae have the following form

$$\hat{A} \leftrightarrow A, \quad \hat{A}\hat{q} \leftrightarrow \left(\alpha q + \beta p + \gamma \frac{\partial}{\partial q} + \delta \frac{\partial}{\partial p}\right)A, \tag{4}$$

where $\alpha, \beta, \gamma, \delta$ are some constants (the correspondancies for the remaining three products can be done analogously).

In [10] the general form of the kerenel $K$ is derived in the class of generalized functions with point supports under the conditions (3) and (4):

$$K(q,p|x,y) = \int_0^1 K_\theta(q,p|x,y) Q(\theta) d\theta,$$

$$K_\theta(q,p|x,y) = \frac{1}{2\pi} \delta(q - \theta x - (1-\theta)y) \exp[ip(x-y)] \tag{5}$$

The integral in (5) is considered as generalized. The generalized function $Q(\theta)$ has the support $[0;1]$ and possesses the properties $\int_0^1 Q(\theta) d\theta = 1$ and $Q(\theta) = Q(1-\theta)$. The last one is a necessary and sufficient condition for the operator $\hat{A}$ to be hermitian. The function $Q(\theta)$ determines the symmetrization rule for non-commuting operators in the product. The kernel $K_\theta$ is said to be partial or a $\theta$-quantization. For example, if $A(q,p) = f(q)p^n$, then the $\theta$-quantization gives the matrix of the operator $\hat{A}_\theta$ of the form

$$\widetilde{A}_\theta(x,y) = (i)^n f(\theta x + (1-\theta)y) \frac{\partial^n}{\partial y^n} \delta(x-y),$$

while the operator itself

$$\hat{A}_\theta = (-i)^n \sum_{k=0}^{n} \binom{n}{k} (1-\theta)^k \left( \frac{\partial^k f(x)}{\partial x^k} \right) \frac{\partial^{n-k}}{\partial x^{n-k}}. \qquad (6)$$

In particular, if $\theta = 0$ we obtain the $pq$-quantization under which all the operators $\hat{p}$ are in left from the operators $\hat{q}$, if $\theta = 1$ we get the $qp$-quantization and $\theta = 1/2$ is corresponding to the Weyl quantization. If one involves the moments $\sigma_k$ of the symmetrization function $Q(\theta)$, then the operator $\hat{A}$ under the quantization with the kernel $K = \int_0^1 K_\theta Q(\theta) d\theta$ takes the form

$$\hat{A} = (-i)^n \sum_{k=0}^{n} \binom{n}{k} \sigma_k \left( \frac{\partial^k f(x)}{\partial x^k} \right) \frac{\partial^{n-k}}{\partial x^{n-k}}, \quad \sigma_k = \int_0^1 \theta^k Q(\theta) d\theta. \qquad (7)$$

If the quantization is hermitian, then the moments of the even (except zero) orders can be arbitrary, while the moments of the odd orders are expressed by means of the requiring formula of the following form

$$\sigma_0 = 1, \quad \sigma_1 = \frac{1}{2}, \quad \sigma_{2k+1} = \frac{1}{2} \left[ 1 + \sum_{n=1}^{2k} (-1)^n \binom{2k+1}{n} \sigma_n \right].$$

In particular, for the Weyl quantization $\sigma_k = 2^{-k}$.

**3. The tomographic representation of the statistical operators**

Let $\hat{\rho}$ be the statistical (density) operator of a quantum system. Consider its matrix $\rho(x,y)$ in the coordiante representation. The mean value of the operator $\hat{A}$ in the state $\hat{\rho}$ can be expressed in the terms of the classical symbol of this operator by the formula

$$\langle \hat{A} \rangle = Tr \hat{A} \hat{\rho} = \int \widetilde{A}(x,y) \rho(y,x) dx dy = \int A(q,p) W(q,p) dq dp, \qquad (8)$$

where a new statistical operator $W(q,p)$ is introduced due to (2) as follows

$$W(q,p) = \int K(q,p|x,y) \rho(y,x) dx dy, \qquad (9)$$

which is a generalization of the Wigner function [11] for the general linear quantization (5). In particular, for the partial quantization $K_\theta$ the Wigner function has the form

$$W_\theta(q,p) = \frac{1}{2\pi} \int e^{ipz} \rho(q - \theta z, q + (1-\theta)z) dz, \quad W(q,p) = \int W_\theta(q,p) Q(\theta) d\theta. \qquad (10)$$

Taking into account that the matrix is hermitian we get from the representation (10) that

$$W_\theta(q,p) = \overline{W}_{1-\theta}(q,p). \tag{11}$$

Thus, under the Weyl quantization $(\theta = 1/2)$ the partial Wigner function is real. In other cases it can not take a place.

For the partial quantization the reverse formula is valid:

$$\rho(x,y) = \frac{1}{2\pi}\int e^{-ip(x-y)} W_\theta(\theta x + (1-\theta)y, p)dp. \tag{12}$$

The symplectic quantum tomogram [1-2] is the positive definite function for the hermitian quantizations defined by the formula

$$f(\xi,\mu,\nu) = Tr(\hat{\rho}\delta(\xi - \mu\hat{q} - \nu\hat{p})) = \frac{1}{2\pi}\int W(q,p)\delta(\xi - \mu q - \nu p)dqdp. \tag{13}$$

The last equation in (13) means that the symplectic quantum tomogram is the Radon transform of the Wigner function. Analogously one can define a tomogram for the partial quantization (it will be not positive definite in general). As for the partial as well as for the general linear quantization there exists a reverse of the formula (13) which allows to reconstruct the Wigner function from the tomogram such that

$$W(q,p) = \frac{1}{2\pi}\int f(\xi,\mu,\nu)e^{i(\xi - \mu q - \nu p)}d\xi d\mu d\nu. \tag{14}$$

The characteristic function of the tomographic distribution is determined by the formula

$$F(s,\mu,\nu) = Tr(\hat{\rho}e^{is(\mu\hat{q}+\nu\hat{p})}) = \int f(\xi,\mu,\nu)e^{is\xi}d\xi =$$
$$= \frac{1}{2\pi}\int W(q,p)e^{is(\mu q + \nu p)}dqdp = \Lambda(s\mu, s\nu), \tag{15}$$

where $\Lambda = \Lambda(k,\omega) = \frac{1}{2\pi}\int W(q,p)e^{i(kq+\omega p)}dqdp$ is the Fourier transform of the Wigner function. Let us show the connection between the Fourier transform of the density matrix and the partial Wigner function:

$$\tilde{\rho}(k,k') = \frac{1}{2\pi}\int \rho(x,y)e^{-i(kx-k'y)}dxdy = \int \Lambda_\theta(k-k',\omega)e^{-i\omega(\theta k' + (1-\theta)k)}d\omega. \tag{16}$$

This formula permits to express the density matrix by means of the characteristic function for the partial tomographic distribution. It is straightforward to check that due to (11) the right hand side of (16) is hermitian although it contains only the Fourier transform of the partial Wigner function.

## 4. The tomographic representation for the states of the driven oscillator

Let us consider the driven oscillator determined by the Hamiltonian [3]

$$H(q,p,t) = \frac{p^2}{2} + \frac{\Omega^2(t)q^2}{2} - \varphi(t)q.  \qquad (17)$$

Under any linear quantization (5) the symbol (17) in the coordinate representation is associated with the operator

$$\hat{H}(t) = -\frac{1}{2}\frac{\partial^2}{\partial x^2} + \frac{\Omega^2(t)x^2}{2} - \varphi(t)x. \qquad (18)$$

The integrals of motion for the Hamiltonian (18) have the form

$$\hat{A}(t) = \frac{i}{\sqrt{2}}\left(\varepsilon(t)\hat{p} - \dot{\varepsilon}(t)x\right) + \delta(t), \quad \hat{A}^+(t) = -\frac{i}{\sqrt{2}}\left(\bar{\varepsilon}(t)\hat{p} - \dot{\bar{\varepsilon}}(t)x\right) + \bar{\delta}(t). \qquad (19)$$

Here $\varepsilon(t)$ is a solution to the equation $\ddot{\varepsilon}(t) + \Omega^2(t)\varepsilon(t) = 0$ under the initial conditions $\varepsilon(0) = 1, \dot{\varepsilon}(0) = i\Omega(0)$, and $\delta(t)$ is a solution to the equation $\dot{\delta}(t) = -\frac{i}{\sqrt{2}}\varepsilon(t)\varphi(t)$ under the initial condition $\delta(0) = 0$. Let us introduce the following notations for the corresponding real values:

$$i(\varepsilon\dot{\bar{\varepsilon}} - \dot{\varepsilon}\bar{\varepsilon}) = 2\alpha(t), \quad \delta\bar{\varepsilon} + \bar{\delta}\varepsilon = 2\beta(t), \quad i\left(\dot{\delta}\bar{\delta} - \delta\dot{\bar{\delta}}\right) = \left(\varepsilon\varphi\bar{\delta} + \bar{\varepsilon}\bar{\varphi}\delta\right)/\sqrt{2} = 2\gamma(t). \qquad (20)$$

Notice that the Wronskian of the system of solutions to the equation $\ddot{\varepsilon}(t) + \Omega^2(t)\varepsilon(t) = 0$ equals $-2i\alpha(t)$ and by means of Liouville formula doesn't depend on time and is equal to $-2i\Omega(0)$. For the sake of shortness let us set $\Omega(0) = 1$. Then, the wave function of the ground state determined by the condition $\hat{A}(t)\psi_0(x,t) = 0$ being normed by one and satisfying the non-stationary Schroedinger equation has the form

$$\psi_0(x,t) = \frac{1}{\left(\pi|\varepsilon|^2\right)^{1/4}}\exp\left(i\frac{\dot{\varepsilon}}{2\varepsilon}x^2 - \frac{\sqrt{2}\delta}{\varepsilon}x - \frac{\beta^2}{|\varepsilon|^2} + i\int_0^t \gamma(\tau)d\tau\right). \qquad (21)$$

If the frequency $\Omega$ doesn't depend on time, then $\varepsilon(t) = e^{i\Omega t}$ and in the absense of the driven forse the formula (21) is transformed into the obvious wave function of the ground state of a harmonic oscillator.

The wave functions of the excited states have the form

$$\psi_n(x,t) = \frac{1}{\sqrt{n!}}\left(\hat{A}^+(t)\right)^n \psi_0(x,t) = \left(\frac{\bar{\varepsilon}}{2|\varepsilon|}\right)^{n/2} \frac{1}{\sqrt{n!}} H_n\left(\frac{x + \sqrt{2}\beta}{|\varepsilon|}\right)\psi_0(x,t), \qquad (22)$$

where $H_n(x)$ are the Hermite polinomials.

The density matrix of a $n$-th excited state of the driven oscillator in the moment of time t is the following

$$\rho_n(x,y) = \frac{1}{2^n n!} H_n\left(\frac{x+\sqrt{2}\beta}{|\varepsilon|}\right) H_n\left(\frac{y+\sqrt{2}\beta}{|\varepsilon|}\right) \psi_0(x,t)\overline{\psi}_0(y,t). \tag{23}$$

For the harmonic oscillator with the Hamiltonian $H(q,p) = q^2/2 + p^2/2$ one should put $|\varepsilon| = 1, \beta = 0, \psi_0(x) = \pi^{-1/4} \exp(-x^2/2)$ in (23). The partial Wigner function (10) is looked too huge even in this simplest case. For example, the ground state leads to the expression

$$W_{\theta,0}(q,p) = \frac{1}{\pi\sqrt{2}\sqrt{\theta^2+(1-\theta)^2}} \exp\left(-q^2 + \frac{1}{2}\frac{((2\theta-1)q+ip)^2}{\theta^2+(1-\theta)^2}\right). \tag{24}$$

Notice that the result of symmetrization of (24) by means of the function $Q(\theta)$ can not be expressed in general through the moments of this function as it was done in formula (7). For certain quantizations the expression for $W(q,p)$ corresponding to states of the harmonic oscillator can be obtained in the evident form. For example, in the case of the Weyl quantization it follows from (24) that for the gound state the Wigner function has the form

$$W_{Weyl,0}(q,p) = \frac{1}{\pi}\exp(-q^2 - p^2),$$

while for the Jordan quantization (a semisum of $pq$- and $qp$-quantizations) we get

$$W_{Jordan,0}(q,p) = \frac{1}{\pi\sqrt{2}}\exp(-q^2/2 - p^2/2)\cos(pq).$$

In virtue of the identity

$$\int_{-\infty}^{\infty} e^{-x^2} H_n(x+a)H_n(x+b)dx = 2^n n!\sqrt{\pi}\,L_n(-2ab), \tag{25}$$

where $L_n(x)$ are the Laguerre polinomials, one can obtain the expression for the Wigner function of the $n$-th excited state of a harmonic oscillator under the Weyl quantization as follows

$$W_{Weyl,n}(q,p) = \frac{(-1)^n}{\pi}\exp(-q^2 - p^2)L_n(2q^2 + 2p^2), \tag{26}$$

Nevertheless it is more convinient to work with the Fourier images of the Wigner function for which the expressions obtained are simple. Thus, by means of the same formula (25) one can derive the expression for the Fourier image of the Wigner function for a harmonic oscillator (we supply it with a top index h) under the linear quantization with the kernel $K_\theta$ determined by formula (5):

$$\Lambda^h_{\theta,n}(k,\omega) = \frac{1}{2\pi}\exp\left(-k^2/4 - \omega^2/4 + ik\omega(\theta - 1/2)\right)L_n\left(k^2/2 + \omega^2/2\right). \tag{27}$$

Now if one introduces the characteristic function $G(s)$ for the function of symmetrization $Q(\theta)$ such that

$$G(s) = \int Q(\theta)e^{is\theta}d\theta, \tag{28}$$

then it is possible to obtain the expression for the Fourier image of the Wigner function under the linear hermitian quantization:

$$\Lambda^h_n(k,\omega) = \int \Lambda^h_{\theta,n}(k,\omega)Q(\theta)d\theta = \frac{G(k\omega)}{2\pi}\exp\left(-k^2/4 - \omega^2/4 - ik\omega/2\right)L_n\left(k^2/2 + \omega^2/2\right). \tag{29}$$

In virtue of (15) the expressions (27), (29) determine the characteristic function of the tomographic distribution for a harmonic oscillator under the linear hermitian quantization (5):

$$F^h_{\theta,n}(s,\mu,\nu) = \frac{1}{2\pi}\exp\left(-\mu^2 s^2/4 - \nu^2 s^2/4 + i\mu\nu s^2(\theta - 1/2)\right)L_n\left(\mu^2 s^2/2 + \nu^2 s^2/2\right)$$

$$F^h_n(s,\mu,\nu) = \frac{G(\mu\nu s^2)}{2\pi}\exp\left(-\mu^2 s^2/4 - \nu^2 s^2/4 - i\mu\nu s^2/2\right)L_n\left(\mu^2 s^2/2 + \nu^2 s^2/2\right) \tag{30}$$

In the same way we derive from (23), (25) the expressions for the Fourier image of the Wigner function for the driven oscillator under the linear partial quantization with the kernel (5):

$$\Lambda_{\theta,n}(k,\omega) = \frac{1}{(2\pi)^2}\exp\left(-\frac{k^2|\varepsilon|^2}{4} - \frac{\omega^2 \varepsilon\dot{\bar\varepsilon}}{2|\varepsilon|^2}\left(i + \frac{\varepsilon\dot{\bar\varepsilon}}{2}\right) + \frac{\sqrt{2}\omega}{|\varepsilon|^2}\left(j\varepsilon\dot{\bar\varepsilon} - \bar\delta\varepsilon\right) + ik\omega\left(\theta - \frac{1}{2}\right)\right) \times$$

$$\times L_n\left(\frac{k^2|\varepsilon|^2}{2} + \frac{\omega^2 \varepsilon\dot{\bar\varepsilon}}{|\varepsilon|^2}\left(i + \frac{\varepsilon\dot{\bar\varepsilon}}{2}\right)\right). \tag{31}$$

Comparing (31) with (27) we see that the fact that the oscillator is driven doesn't affect on the quantization because the dependence on $\theta$ of the Fourier image of the Wigner fuction doesn't contain the master parameters $\varepsilon$ and $\delta$. This is the important conclusion which shows that under the external influence on the quantum system considered its quantization is not changed. The expression for the corresponding characteristic function is obtained from (31) under taking into account (15) and as well as for (29) and (30) is almost the same. So we shall not write out it. Also we shall not write out the simmetrized hermitian Wigner function because this expression will differ from (31) as well as in (29) only by the factor. Instead of $\exp(ik\omega(\theta - 1/2))$ it will appear $G(k\omega)\exp(-ik\omega/2)$.

## 5. Evolution equation for a tomographic distribution

Taking into account the Liouville-von Neumann equation for the density operator

$$i\partial_t \hat{\rho} = [\hat{H}, \hat{\rho}], \quad (32)$$

one can derive the evolution equations for the Wigner function as well as for the tomographic distributions. For the Weyl quantization the evolution equation of the Wigner function was obtained by Moyal in [12]. The evolution of the tomographic distributions of a driven oscillator for this quantization was derived in [3, 4]. For an arbitrary linear quantization of the Hamiltonian $H(q,p)$ the evolution equation of the Wigner function was obtained by one of the authors in [10]. In particular, the evolution equation of the partial Wigner function has the form

$$i\partial_t W_\theta(q,p) = \int h(k,\omega) G(k\omega) \exp(i(kq + \omega p + (1-\theta)k\omega)) \Delta_\theta W_\theta \, dk \, d\omega, \quad (33)$$

where $h(k,\omega) = (2\pi)^{-2} \int H(q,p) e^{-ikq - i\omega p} dq \, dp$ is the Fourier image of the classical Hamiltonian $H(q,p)$ and the function $G(s)$ was defined above in (28). To breve the notation in (33) we introduced the difference operator

$$\Delta_\theta W_\theta = W_\theta(q - \theta\omega, p + (1-\theta)k) - W_\theta(q + (1-\theta)\omega, p - \theta k). \quad (34)$$

Notice that the dependence on the parameter of a quantization $\theta$ is included in (34) by means of two aspects: in the partial Wigner function itself as it was in Examples (24) and (27) as well as in its arguments shifted in the corresponding way. The Hamiltonian $\hat{H}$ is quantized by the full hermitian quantization such that in (33) the factor $G(k\omega)$ has appeared. The equation for the full Wigner function follows from (33) by integrating it in $\theta$ with the symmetrization function $Q(\theta)$:

$$i\partial_t W(q,p) = \int h(k,\omega) G(k\omega) \exp(i(kq + \omega p + k\omega)) dk \, d\omega \int_0^1 Q(\theta) \exp(-i\theta k\omega) \Delta_\theta W_\theta \, d\theta. \quad (35)$$

The equations obtained allow to recognize the structure of the integro-difference operator determining the evolution of a quasi-probability. The important peculiarity of Equation (35) is that the evolution of the full Wigner function can not be represented as a certain operator acting to it directly as it takes place for the partial Wigner function $W_\theta$. The reason is the factor $\exp(-i\theta k\omega) G(k\omega)$ which equals to one only for the Weyl quantization. The same factor resists to carrying the action of the difference operator $\Delta_\theta$ from the Wigner function to the Hamiltonian as it can be done for the Weyl quantization with $\theta = 1/2$ (in this form the Wigner equation is represented more frequently):

$$i\partial_t W(q,p) = \int \Lambda(k,\omega)\exp(ikq + i\omega p)\left[H\left(q - \frac{\omega}{2}, p + \frac{k}{2}\right) - H\left(q + \frac{\omega}{2}, p - \frac{k}{2}\right)\right]dkd\omega. \quad (36)$$

Nevertheless if the Hamiltonian is represented as a sum of the form

$$H(q,p) = E(p) + \Phi(q), \quad (37)$$

then such a transfomation of the evolution equation (33) can be fulfilled:

$$i\partial_t W_\theta(q,p) = \int dkd\omega \Lambda_\theta(k,\omega)\exp(ikq + i\omega p)\Delta_\theta H, \quad (38)$$

$$\Delta_\theta H = E(p + (1-\theta)k) - E(p - \theta k) + \Phi(q - \theta\omega) - \Phi(q + (1-\theta)\omega).$$

It follows from (38) that it is convinient to write the evolution equation for the Wigner function in terms of the Fourier images. Denote $\widetilde{\Lambda}_\theta(k,\omega) = \Lambda_\theta(k,\omega)\exp(-ik\omega(1-\theta))$ and as in (33) marking by lower indeces the Fourier images of the Hamiltonian we get

$$i\partial_t \widetilde{\Lambda}_\theta(k,\omega) = \int \left(1 - e^{ik(\omega-\omega')}\right)\left(\widetilde{\Lambda}_\theta(k,\omega')E_{\omega-\omega'} - \widetilde{\Lambda}_\theta(\omega',k)\Phi_{\omega-\omega'}\right)d\omega'. \quad (39)$$

The same equation (39) determines due to (15) the evolution equation of the characteristic function of a tomographic distribution.

If $E(p) = p^2/2$, then Equation (38) takes the form

$$\partial_t W_\theta(q,p) + p\nabla_q W_\theta(q,p) - \frac{i}{2}(1-2\theta)\nabla_q^2 W_\theta(q,p) =$$
$$= -i\int W_\theta(q,s)e^{i\omega(p-s)}\left[\Phi(q-\theta\omega) - \Phi(q+(1-\theta)\omega)\right]\frac{dsd\omega}{2\pi}. \quad (40)$$

In particular, for the Hamiltonian (17) we get

$$\partial_t W_\theta(q,p) + \left(p\nabla_q - (\varphi - \Omega^2 q)\nabla_p\right)W_\theta(q,p) = \frac{i}{2}(1-2\theta)\left(\nabla_q^2 - \Omega^2\nabla_p^2\right)W_\theta(q,p). \quad (41)$$

Equation (41) in the Fourier images has the form

$$\partial_t \Lambda_\theta(\mu,\nu) + \left(u\partial_\nu + (i\varphi + \Omega^2\partial_\mu)\right)\Lambda_\theta(\mu,\nu) = \frac{i}{2}(1-2\theta)\left(\Omega^2\nu^2 - \mu^2\right)\Lambda_\theta(\mu,\nu). \quad (42)$$

Using (15) one can find the evolution equation for the characteristic function:

$$\partial_t F_\theta(s,\mu,\nu) = \partial_t \Lambda_\theta(s\mu,s\nu) =$$
$$= -\left(u\partial_\nu + (is\varphi + \Omega^2\partial_\mu)\right)F_\theta(s,\mu,\nu) + \frac{i}{2}(1-2\theta)s^2\left(\Omega^2\nu^2 - \mu^2\right)F_\theta(s,\mu,\nu) \quad (43)$$

The evolution equation for the quantum tomogram of a driven oscillator is obtained from (13) and (42) such that

$$\partial_t f_\theta(\xi,\mu,\nu) = \frac{1}{2\pi}\int \exp(ik\xi)\partial_t \Lambda_\theta(\mu k,\nu k) =$$
$$= -\left(u\partial_\nu + (\varphi\partial_\xi + \Omega^2\partial_\mu)\right)f_\theta(\xi,\mu,\nu) - \frac{i}{2}(1-2\theta)\left(\Omega^2\nu^2 - \mu^2\right)\frac{\partial^2}{\partial\xi^2}f_\theta(\xi,\mu,\nu) \quad (44)$$

The evolution equation for the full Wigner function is derived from (41) by integrating this equality multiplied by the symmetrization function in $\theta$ according to (10). If the quantization is hermitian, then as a result the left part of (41) becomes real. The right part of (41) becomes real as well if one takes into account the relation in (11) joint with the condition $Q(\theta) = Q(1-\theta)$ which is neccessary for a quantization to be hermitian. Thus we obtain

$$\partial_t W(q,p) + \left(p\nabla_q - (\varphi - \Omega^2 q)\nabla_p\right)W(q,p) = \frac{i}{2}\left(\nabla_q^2 - \Omega^2 \nabla_p^2\right)\int_0^1 (1-2\theta)Q(\theta)W_\theta(q,p)d\theta. \quad (45)$$

The right hand side of Equation (45) is the quantum effect appeared as a consequence of taking into account the dependence of the Wigner function on the rule of a quantization. The distinction of the right hand side from zero for the hermitian quantizations which do not coincide with the Weyl quantization can be used for determining how the quantum dynamical system was quantized experimentally.

Analogously, due to (44) we obtain the evolution equation for the quantum tomogram in the case of the full hermitian quantization:

$$\partial_t f(\xi,\mu,\nu) + \left(\mu\partial_\nu + (\varphi\partial_\xi + \Omega^2 \partial_\mu)\right)f(\xi,\mu,\nu) =$$
$$= -\frac{i}{2}\left(\Omega^2\nu^2 - \mu^2\right)\frac{\partial^2}{\partial\xi^2}\int Q(\theta)(1-2\theta)f_\theta(\xi,\mu,\nu)d\theta. \quad (46)$$

## 6. Conclusion

We derived the evolution equations for the Wigner function (45) as well as for the symplectic quantum tomogram (46) of the driven quantum oscillator in the case of the general linear quantization. Both the equations have the non-trivial integral part which is not included into the equation for the case of the Weyl quantization. The presence of such a term could allow to check experimentaly how the system was quantized.

**Acknowledgments**

The work is partially supported by INTAS grant Ref. Nr. 06-1000014-6077 (G.G. Amosov) and by the Russian Foundation for Basic Research under Project Nr. 07-02-00598 (G.G. Amosov and V.I. Man'ko).